\documentclass[12pt,preprint]{aastex}
\usepackage{amsmath,amssymb,graphicx,natbib}

\newsavebox{\astrutbox}
\sbox{\astrutbox}{\rule[-5pt]{0pt}{20pt}}

\newcommand{\gapprox}{\lower.4ex\hbox{$\;\buildrel >\over{\scriptstyle\sim}\;$}}
\newcommand{\lapprox}{\lower.4ex\hbox{$\;\buildrel <\over{\scriptstyle\sim}\;$}}

\bibpunct[; ] {(}{)}{,}{}{}{,}
\shorttitle{Magetorotational turbulence in stratified boxes} 
\shortauthors{Bodo, Cattaneo, Mignone \& Rossi} 

\begin{document} 
 
\title{Fully Convective Magnetorotational Turbulence in Stratified Shearing Boxes} 
 
 \author{G. Bodo\altaffilmark{1},        
         F. Cattaneo\altaffilmark{2}, 
         A. Mignone\altaffilmark{3},
         P. Rossi\altaffilmark{1} }
  
 \altaffiltext{1}{INAF, Osservatorio Astronomico di Torino, Strada Osservatorio 20, Pino Torinese, Italy}
 
 \altaffiltext{2}{The Computation Institute, The University of Chicago, 
              5735 S. Ellis Avenue, Chicago IL 60637, USA}

 \altaffiltext{3}{Dipartimento di Fisica Generale, Univresit\'a di Torino, via Pietro Giuria 1, 10125 Torino, Italy} 
 
\begin{abstract} 
We present a numerical study of turbulence and dynamo action in stratified shearing boxes with zero magnetic flux. We assume that the fluid obeys the perfect gas law and has finite (constant) thermal diffusivity. 
We choose radiative boundary conditions at the vertical boundaries in which the heat flux is proportional to the fourth power of the temperature. We compare the results with the corresponding cases in which fixed temperature boundary conditions are applied. The most notable result is that the formation of a fully convective state in which the density is nearly constant as a function of height and the heat is transported to the upper and lower boundaries by overturning motions is robust and persists even in cases with radiative boundary conditions. Interestingly, in the convective regime, although the diffusive transport is negligible the mean stratification does not relax to an adiabatic state.

\end{abstract} 

\section{Introduction}
What determines the internal structure of accretion discs and how is angular momentum transported in these objects are two fundamental problems in plasma astrophysics. It is universally believed that magnetic fields play an important role in the mechanism of disc destabilization and in the generation of turbulence. More specifically, the magneto-rotational instability (MRI) has been invoked as the most likely explanation for the origin of disc turbulence, and indeed, it provides an elegant framework for the discussion of such issues \citep{Balbus91}. In those cases in which there is no net magnetic flux, the magnetic field necessary to drive the MRI must be re-generated by the turbulent motions. The problem of the origin of disc turbulence then becomes one of driving a turbulent dynamo in an accretion flow.  Because of the inherently nonlinear nature of this problem--dynamo action must set in as a subcritical nonlinear instability--most of what is currently known about disc dynamos is based on numerical studies. The most commonly adopted configuration for such studies is that of a local, Cartesian representation known as the shearing-box approximation \citep{Hawley95}. Recently, many such studies have included some form of vertical stratification, typically described as an isothermal atmosphere \citep{Davis10, Shi10, Oishi11}. For these cases the dynamo is highly inhomogeneous, with strikingly different behaviors in the dense regions near the equatorial plane and in the more tenuous overlying layers \citep{Gressel10, Guan11, Simon12}. 
In a recent paper, Bodo and collaborators \citep{Bodo12}--hereinafter Paper I--have considered cases with  stratified shearing-boxes with an ideal equation of state, dissipative internal heating and a 
simplified treatment of heat transport in terms of radiative diffusion. Their results showed that
if the thermal conduction were efficient, the solutions resembled the isothermal cases,
if it were not, then a convective state would set in that dramatically altered the vertical structure of the disc. In this new fully convective state, the density was nearly constant across the disc and a radically different type of dynamo action became operative. Significantly, one that was characterized by the production of substantial amounts of toroidal flux and  more efficient angular momentum transport relative to the isothermal cases. In their work, \citet{Bodo12} used the isothermal atmosphere as initial condition and imposed thermal boundary conditions in which the temperature was fixed--and equal to the initial value; this was done so that a direct comparison could be made with the isothermal cases. 
Because the appearance of a fully convective state with an associated efficient dynamo action is novel and can have important consequences for our understanding of accretion flows, it is necessary to ensure that the fully convective states persist when less restrictive boundary conditions are used. We address this issue here and consider instead systems with radiative boundary conditions, i.e. ones in which the heat flux is proportional to some power of the temperature--for instance, the fourth power for black body radiation. These conditions allow the overall temperature of the layer to ``slide" and choose its own self-consistent value--possibly one very different from the initial one. We also note, that the radiative boundary conditions take care of another peculiarity of the fixed temperature case, namely the invariance with respect to the rescaling of the units of mass. If  radiative conditions are imposed, the constant of proportionality between the flux and whichever power of the temperature is chosen, uniquely fixes the unit of mass in terms of the other disc properties. 

\section{Formulation} \label{formulation} 
The current formulation is essentially the same as in Paper I. We perform three-dimensional, numerical simulations of a perfect gas with finite thermal conduction in a shearing box with vertical gravity.  A detailed presentation of the shearing box approximation and the relevant equations can be found in \citet{Hawley95}.  The computational domain covers the region $1 \times \pi \times 6$,  where our unit of length is the pressure scale height in the initial isothermal state.   In the vertical direction the box is symmetric with respect to the equatorial plane $z = 0$,  where gravity changes sign. We assume periodic  boundary conditions in the $y$ direction and shear periodic conditions in the $x$ direction.  In the vertical direction,  we assume that  the upper and lower boundaries ($z = \pm 3$), are impenetrable, stress free and in hydrostatic balance, the magnetic field is taken purely vertical. We consider an optically thick plasma and approximate the radiative transport by a diffusion process which we model by a thermal conduction term, with thermal diffusivity $\kappa$,  in the energy equation. In general  $\kappa$ depends on density and temperature. However,  a more realistic  treatment that incorporates the correct dependencies on density and temperature would have an effect in the conductive regimes, but hardly any in the convective regimes in which the energy transport is all by advection and the contributions by conduction are all but negligible. Since we are primarily interested in the convective solutions, and use the conductive ones mostly as reference states, we keep the same formulation as in Paper I and assume a constant $\kappa$.
In contrast to Paper I,  here, we adopt  a thermal radiative boundary condition of the form:
\begin{equation}
\frac{dT}{dz} \pm \frac{\Sigma}{ \kappa \rho} T^4 =0  \qquad {\rm at}\qquad z=\pm 3,
\label{thermal_BC}
\end{equation}
where $\Sigma$ is a dimensionless quantity proportional to Stefan-Boltzmann constant. Its precise definition and physical meaning will be discussed later, for now it should be considered simply as a constant defining the thermal boundary conditions. 

Some care is necessary in the choice of the initial conditions. In Paper I, each simulation began from a state of isothermal hydrostatic balance. Clearly, such  state does not satisfy the thermal boundary conditions (\ref{thermal_BC}). One could ignore this fact and use it anyway, relying on the ability of the code to impose the boundary conditions after the first time-step---and for all subsequent steps. There are at least two reasons why this approach is not a good idea.  
Numerically, we found that starting from an isothermal state inevitably lead to the formation of very sharp unresolved boundary layers and to the breakdown of the numerical procedure within a few tens of steps.
In principle one could overcome this problem by purely numerical artifices, like adopting an initial grid with high resolution near the boundaries, or greatly reducing the time-steps at the beginning of the calculations. These procedures notwithstanding, there is another, more physical reason why things may not work out anyway.
It should be noted that there is an isothermal solution satisfying (\ref{thermal_BC}); it is the one with zero temperature. If the layer is already in a turbulent state the dissipative heating keeps the solution away from this singular case. If, on the other hand, the turbulent state has not yet evolved, because the MRI has not had time fully to develop, the overall solution could be attracted to the zero temperature solution. Even in cases in which the descent to zero temperature is interrupted by the development of the MRI, the system may, by the time the MRI heating kicks in,  have evolved to a state that is thermodynamically very far from the eventual stationary state. This possibility could, in principle, eventually yield a sensible answer but would require a long integration time to reach a stationary state.  
Instead we use different approach that takes care of both problems. We use the stationary state solutions of Paper I to start the calculations with the new boundary conditions.  
In the new calculations, the value of  $\Sigma$ is chosen so that condition (\ref{thermal_BC}) is approximately satisfied by the averages over the vertical boundaries. With this setup, the boundary condition (\ref{thermal_BC}) is not satisfied initially in a pointwise manner but, at least, it is satisfied in the mean. This way, the initial conditions are already ``near" a stationary state with fully developed MRI driven turbulence.

In the present communication we consider three representative cases defined by their thermal  diffusivity which takes the values of $\kappa=1.2 \times 10^{-1}, 2 \times 10^{-2}, 4 \times 10^{-4}$. In the parlance of Paper I, these values correspond to cases in the conductive regime, near the critical value, and in the convective regime, respectively.  In all three cases $\Sigma=0.1$.  We adopt a resolution that is twice that of Paper I, namely $64 \times 192 \times 384$. 
All  simulations are carried out with the PLUTO code, with a second order accurate scheme, HLLD Riemann solver. The thermal conduction is treated explicitly for small values of $\kappa$, and by super-time stepping for larger ones \citep [see][]{Mignone07}.

\section{Results} \label{results} 
\begin{figure}[htbp]
   \centering
   \includegraphics[width=8cm]{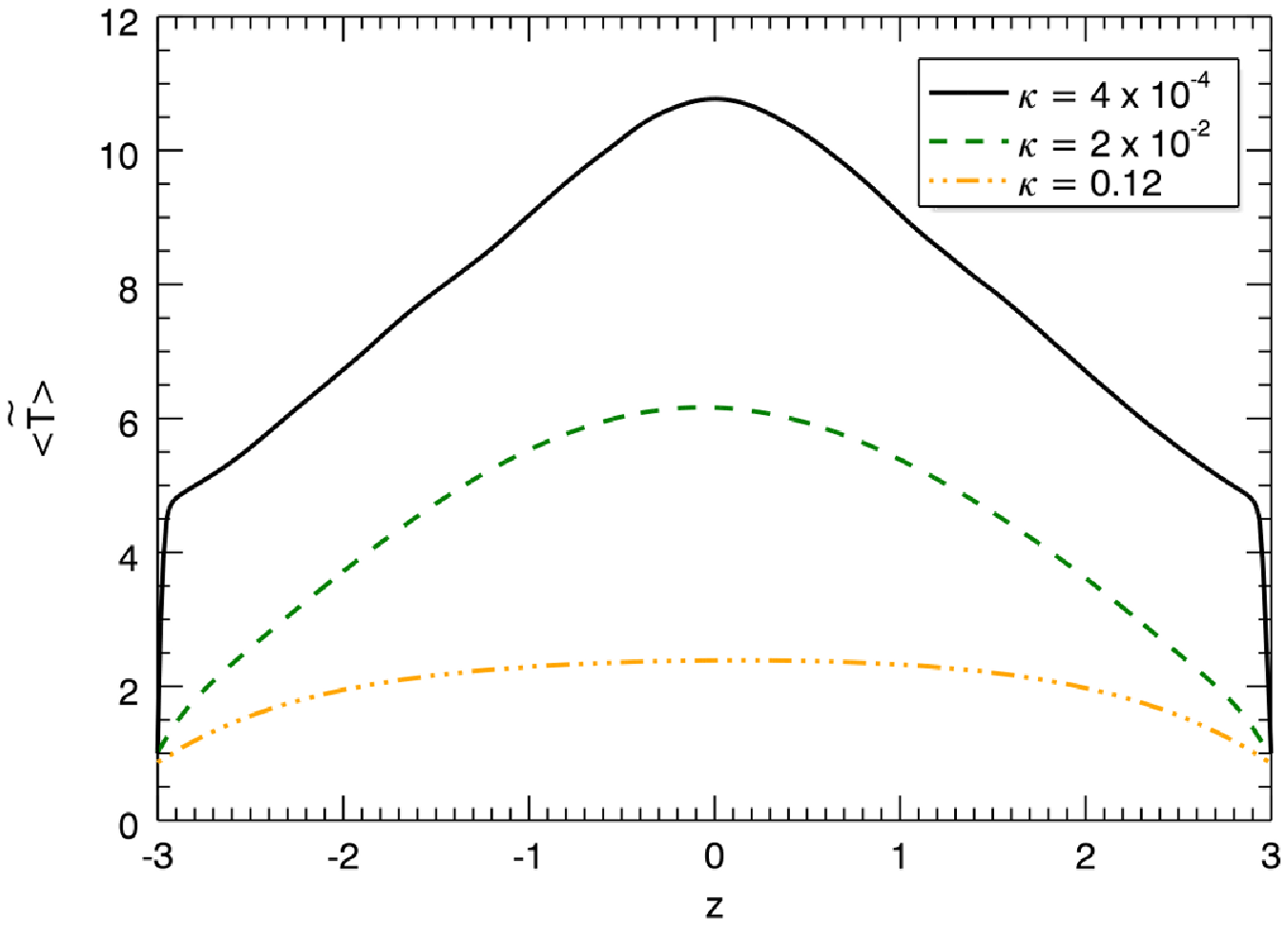} 
   \includegraphics[width=8cm]{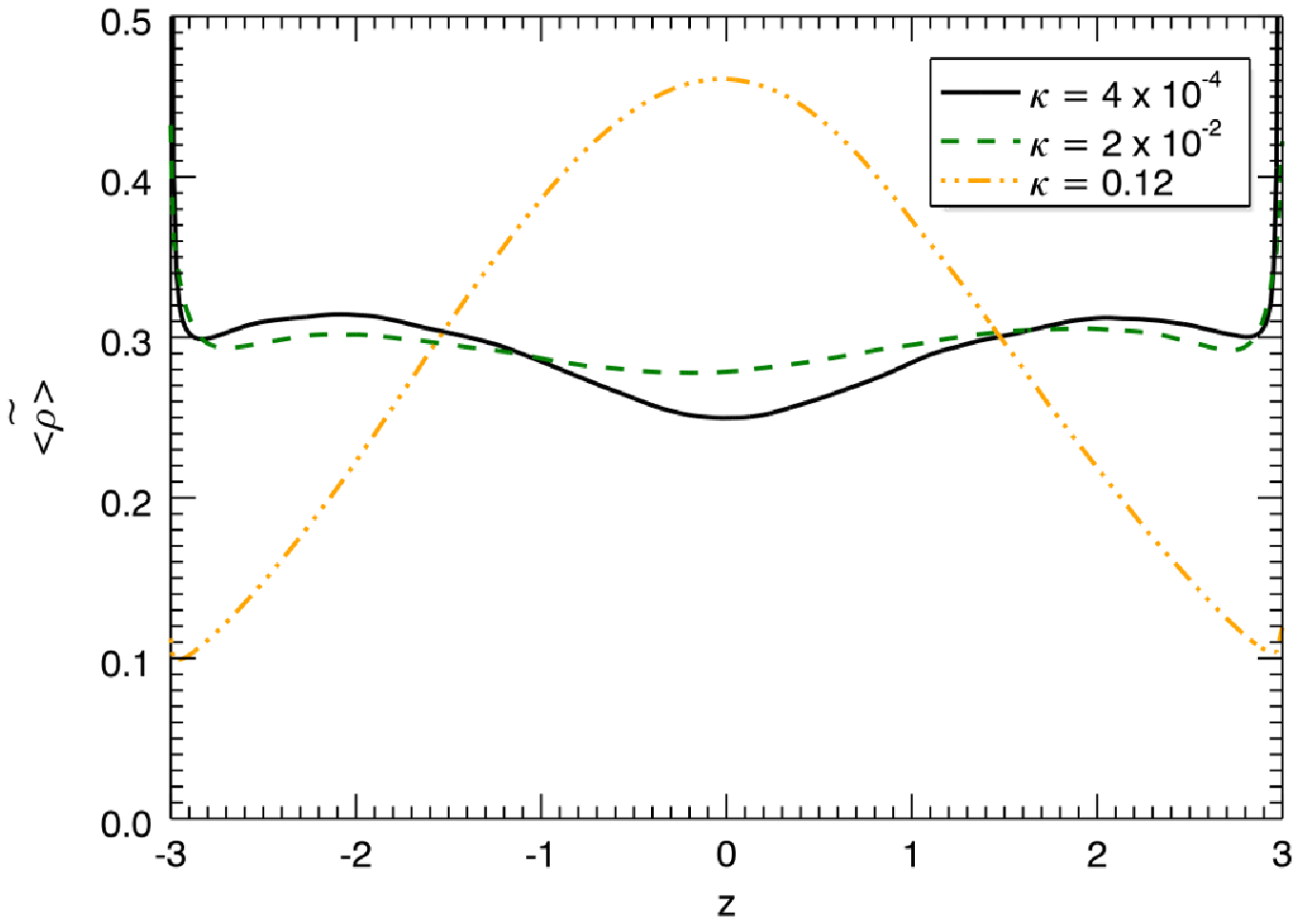} 
   \includegraphics[width=8cm]{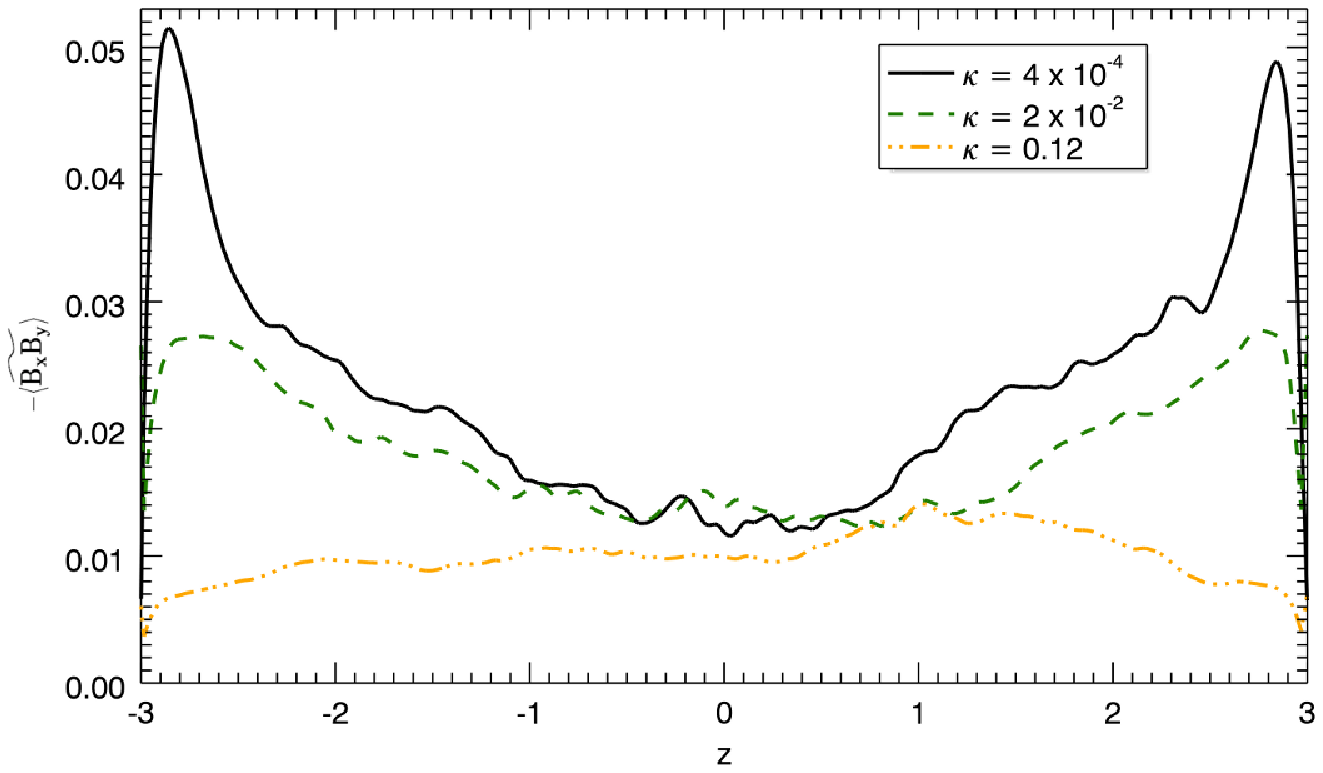} 
   \caption{Horizontally and time averaged profiles of the temperature  density and Maxwell stresses as functions of $z$. The three cases correspond to values of $\kappa$ equal to $4 \times 10^{-4}$
(solid, black lines), $2 \times 10^{-2}$ (dashed, green lines), and $1.2 \times 10^{-1}$ (dash-dotted, yellow lines) respectively.
 }
   \label{fig:temp-rho-max}
\end{figure}

\begin{figure}[htbp]
   \centering
   \includegraphics[width=8cm]{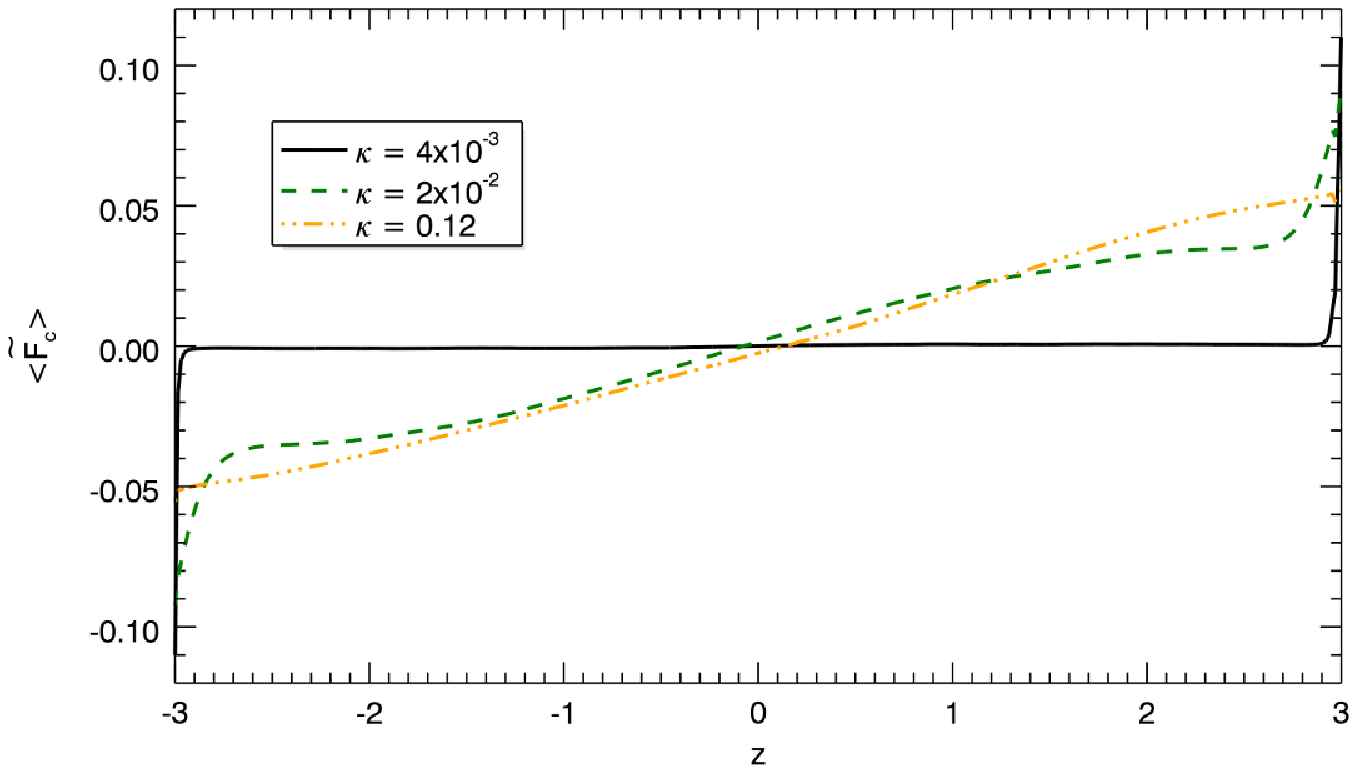} 
   \includegraphics[width=8cm]{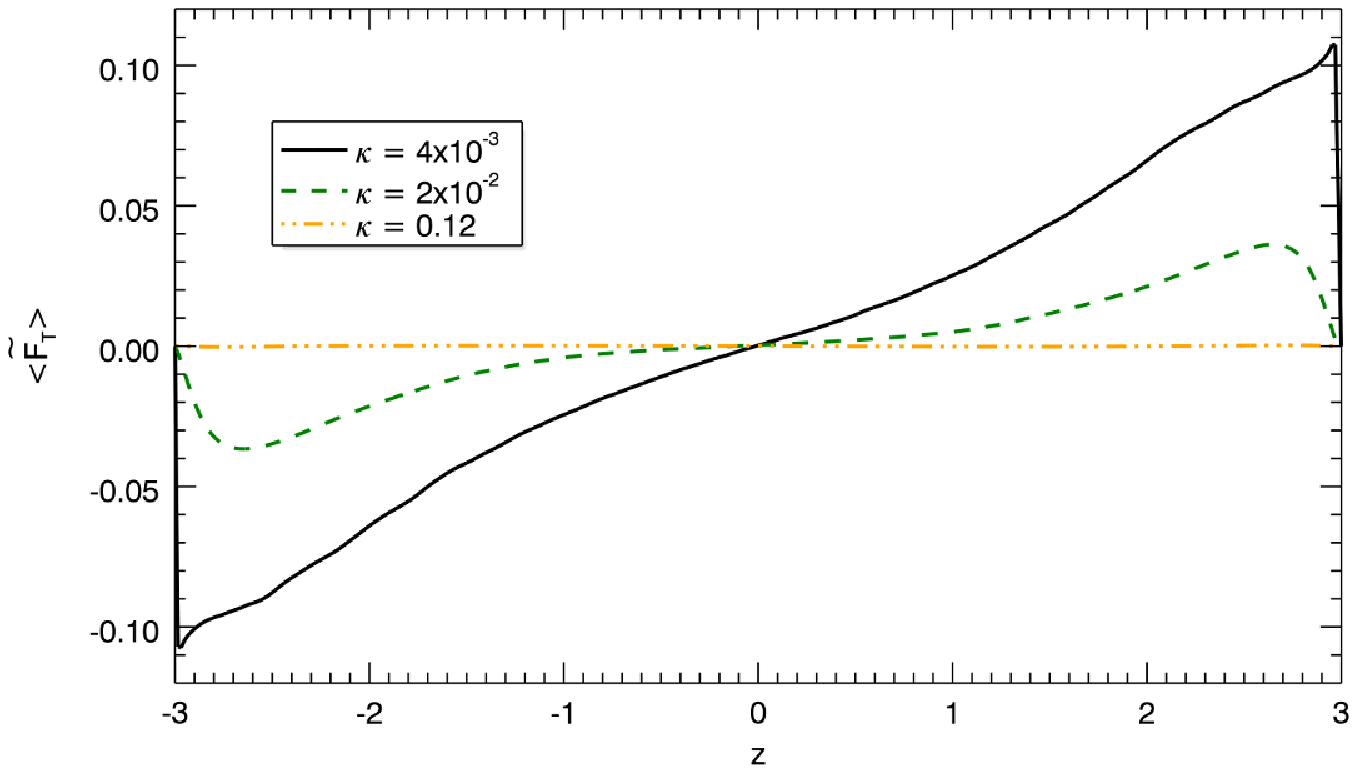} 

   \caption{Horizontally and time averaged profiles of the conductive flux $F_c$, and convective flux $F_T$ as functions of $z$. The three cases correspond to values of $\kappa$ equal to $4 \times 10^{-4}$
(solid, black lines), $2 \times 10^{-2}$ (dashed, green lines), and $1.2 \times 10^{-1}$ (dash-dotted, yellow lines) respectively.
 }
   \label{fig:temp-flux}
\end{figure}

\begin{figure}[htbp]
   \centering

   \includegraphics[width=10cm]{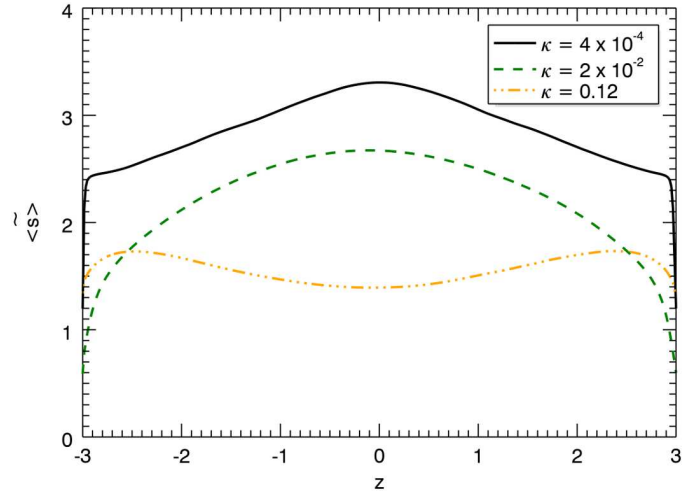} 

   \caption{Horizontally and time averaged profiles of the entropy as a function of $z$. The three cases correspond to values of $\kappa$ equal to $4 \times 10^{-4}$
(solid, black lines), $2 \times 10^{-2}$ (dashed, green lines), and $1.2 \times 10^{-1}$ (dash-dotted, yellow lines) respectively.
 }
   \label{fig:entropy}
\end{figure}

\begin{figure}[htbp]
   \centering
   \includegraphics[width=8cm]{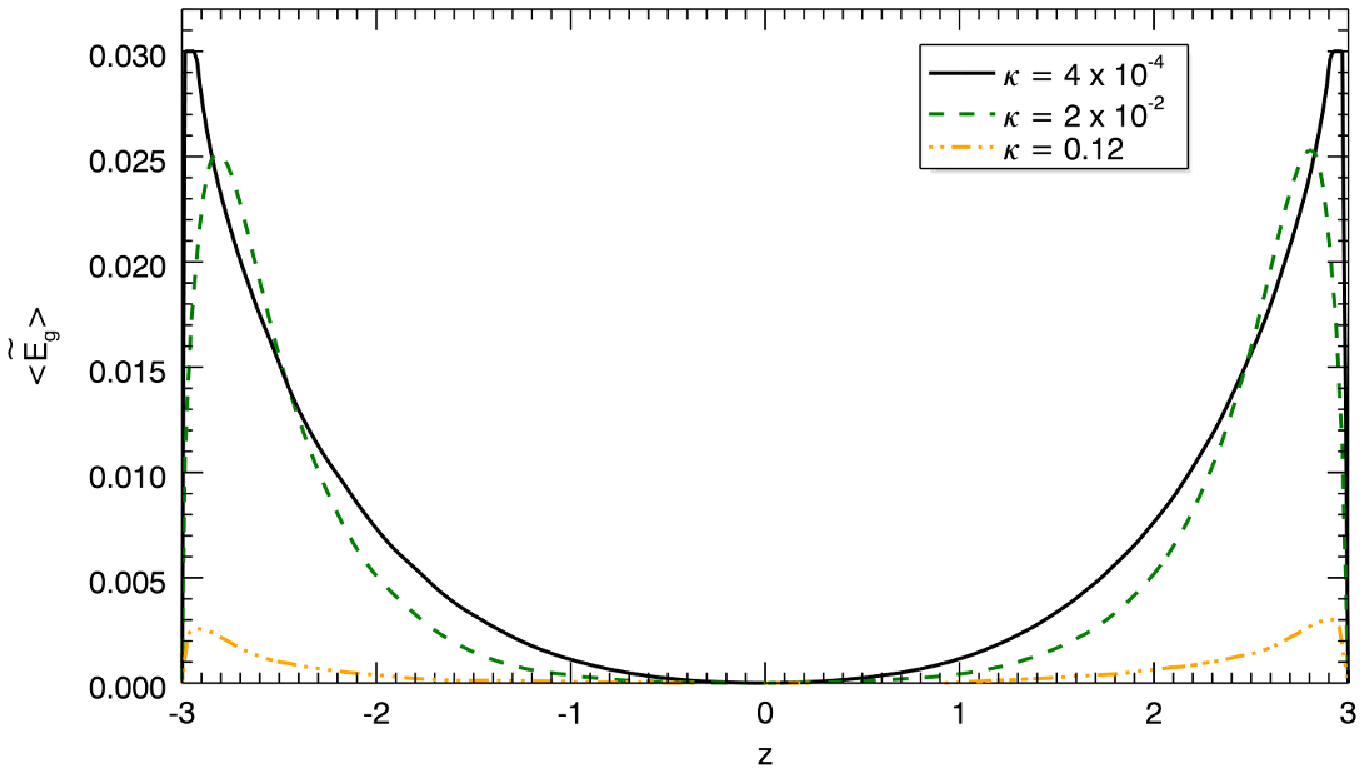} 
   \includegraphics[width=8cm]{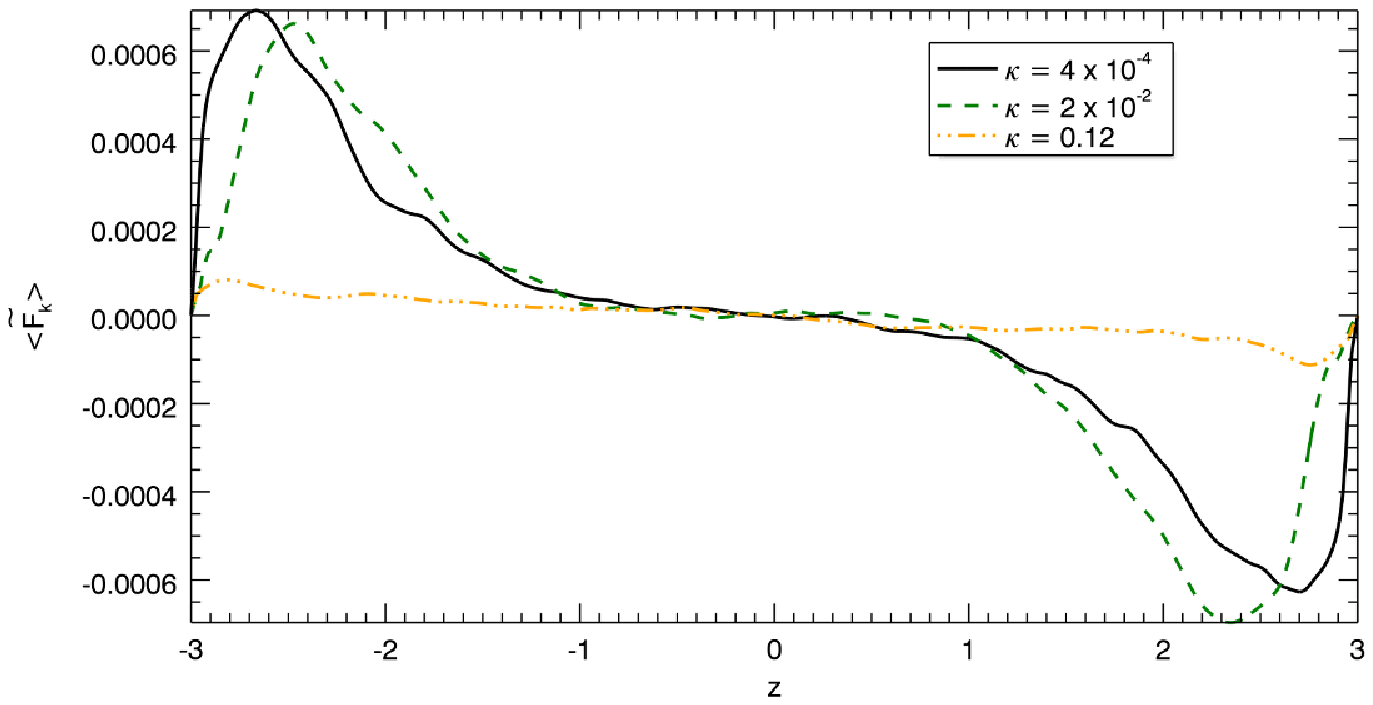} 

   \caption{Horizontally and time averaged profiles of the rate of buoyancy work $E_s$, and vertical kinetic energy flux $F_k$ as functions of $z$. The three cases correspond to values of $\kappa$ equal to $4 \times 10^{-4}$
(solid, black lines), $2 \times 10^{-2}$ (dashed, green lines), and $1.2 \times 10^{-1}$ (dash-dotted, yellow lines) respectively.
 }
   \label{fig:buoy-kin}
\end{figure}

We now describe the properties of the stationary states for the three cases defined above. The typical integration  extends over 1000  time units (approximately 160 rotations) which include both a rapid relaxation phase lasting a few orbits and a stationary state. Figure \ref{fig:temp-rho-max} shows the averages on horizontal planes, and in time over the entire simulation, of the temperature, density and Maxwell stresses for all three cases. Comparing these curves with the corresponding ones
in Paper I shows that the overall qualitative features of the solutions remain largely unchanged. 
In particular, as $\kappa$ decreases, and the solutions move from the conductive to the convective regime,  the temperature profile changes from approximately parabolic to the ``tent" profile,  the density changes from near Gaussian to nearly constant, and the Maxwell stresses increase and become more  concentrated near the boundaries. 
Although these results are
not entirely surprising, specially in view of our strategy for choosing $\Sigma$, it is important to note that  they do show that the stationary states are stable --i.e. there is no tendency for the layers to drift to a completely different  temperature--by no means a foregone conclusion in such a strongly nonlinear system. 
The transition from a conductive to a convective regime is also apparent in the nature of the thermal transport. Figure \ref{fig:temp-flux} shows the conductive and convective  fluxes, as defined in Paper I,  as  functions of $z$ for the three cases. Except inside the two  very thin boundary layers, the conductive flux for the case with the smallest value of $\kappa$ is all but zero, and all of the flux is carried by convection throughout the bulk of the fluid. Interestingly, even though, in this sense, the convection is extremely efficient, it does not lead to an adiabatic stratification. In fact, as shown in Figure \ref{fig:entropy}, the entropy profile remains superadiabatic even when convection is efficient, in contrast with regular thermal convection in which the average stratification becomes very close to adiabatic.  In view of this difference, it is natural to ask to what extent we should think of the motions that are responsible for the vertical energy flux here as convective. Some insight into this issue is provided by the curves in Figure \ref{fig:buoy-kin} showing the horizontally averaged profiles of the rate of buoyancy work and vertical kinetic energy flux defined, respectively, by

\begin{equation}
E_g=g(z)\rho v_z, \qquad  F_k = \frac{1}{2} \rho v_z^3.
\label{eq:work-flux}
\end{equation}
Here $g(z)$ is the vertical component of the gravitational acceleration given, in a shearing-box, by $g(z) = -\Omega^2 z$. Clearly $E_g$ shows that the convective motions are indeed convective, in the sense that they are driven by buoyancy forces. The rate of doing work is close to negligible in the conductive case, while in the convective cases it becomes large and strongly peaked near the boundaries where gravity is largest and radiative cooling is most effective. The kinetic energy flux also has the familiar shape characteristic of thermally driven convection with a downward directed flux indicative of an asymmetry between upflows and downflows, with  strong concentrated downflows and gentler broader upflows \citep{hurtoom84}. 
Finally, and for completeness, we remark that the magnetic field structure in the current simulations is similar to that in the corresponding simulations in Paper I. 

\section{Conclusions} \label{conclusions}
The purpose of the current work was to extend the solutions of Paper I to the radiative boundary conditions defined by (\ref{thermal_BC}). Our main result has been to find that the qualitative features of the solutions remained largely unchanged. Since our choice of $\Sigma$ was was such that the old solutions almost satisfied the new boundary conditions, this result is perhaps not too surprising. Nevertheless, it behooves us to speculate what would happen had we chosen a dramatically different value. Also, we should clarify what is the physical meaning associated with  $\Sigma$
and its exact relationship to Stefan-Boltzmann constant. 

A simple argument can be made to suggest that the effects of varying $\Sigma$ should be mostly to adjust the overall working temperature of the layer relative to the boundary value, but not to change the shape of the temperature profile. This can be seen by considering a idealized diffusion equation with internal heating
\begin{equation}
\frac{\partial \theta}{\partial t} = \kappa \frac{\partial^2 \theta}{\partial z^2} +h,
\label{eqn:diffusion}
\end{equation}
where $h$ is a (spatially) uniform heating rate and, for simplicity, we adopt linearized radiative conditions
of the form
\begin{equation}
\frac{\partial \theta}{\partial z} \pm \frac{\Sigma}{\eta} \theta= 0 \quad{\rm at} \quad z=\pm1,
\label{bc:simple}
\end{equation}
where$\eta =\rho \kappa/ 4 T_o^3$.
In a steady state the solution is given by 
\begin{equation}
\theta(z) = -\frac{h}{2\eta}z^2 + \theta_0, \quad {\rm where} 
\quad \theta_0=h ( \frac{1}{2\eta} + \frac{1}{\Sigma}).
\label{eqn:sol}
\end{equation}
Clearly, the profile depends on a balance between the heating rate $h$ and the heat transport coefficient $\eta \propto \kappa$ but not on $\Sigma$. The latter only determines the overall temperature offset. Furthermore, the smaller $\Sigma$, the higher the temperature. 

To understand how $\Sigma$ relates to Stefan-Boltzmann constant and, in fact, how it determines the unit of density we need to consider the dimensionless form of the disc equations in the shearing-box approximation. For definiteness we assume that $C_s$ is the adiabatic sound speed, and choose $\Omega^{-1}$ and $C_s/\Omega$ as the units of time and distance respectively. With this choice, the continuity equation has no adjustable coefficients, while the equations of conservation of momentum and energy and the induction equation have three dimensionless numbers appearing in front of the dissipative terms and proportional respectively to the inverses of the Reynolds, Peclet, and magnetic Reynolds numbers. We note that the dimensionless equation of state for a perfect gas also has no adjustable coefficients. If one adopts fixed temperature boundary conditions, as was the case in Paper I, this fixes the sound speed, and hence the size of the disc, but not its density; the mass within the disc is undefined. If, on the other hand, one adopts, black-body radiative conditions, as we did here, the dimensionless boundary conditions become (\ref{thermal_BC}) with $\kappa$  the inverse Peclet number, and 
\begin{equation}
\Sigma^2 = \left[\frac{\sigma^2}{{\cal R}^3} \frac{(\gamma -1)^2}{\gamma^3}\right] \frac{T_0^5 \mu}{\rho_0^2},
\label{def_sigma}
\end{equation}
where, $\sigma$ is Stefan-Boltzmann constant, ${\cal R}$ is the ideal gas constant, $\gamma$ is the ratio of specific heats, and $\mu$ is the mean molecular weight. Although not immediately apparent, $\Sigma$ is proportional to the ratio of two energy fluxes: the flux of black body radiation at temperature $T_0$, and  the kinetic energy flux  of a fluid moving at speed $C_s$. In (\ref{def_sigma}) the quantities in the square brackets are fixed physical constants, thus, clearly, changing $\Sigma$ at fixed temperature uniquely defines the unit of density (mass). 

We now briefly discuss the issue of what gets homogenized in the convective regime. All the indicators are that the vertical energy transport is by thermally driven convection. Thus, at face value, one would expect that the entropy should be homogenized and that the average stratification should be nearly adiabatic. As shown in Figure \ref{fig:entropy} this is not the case; the layer remains superadiabatic and it is the density instead that becomes homogenized. Thus the convection is efficient in the sense that it carries practically all of the heat, but inefficient because it does not relax the layer to an adiabatic state. Superficially this may appear strange, but actually it is not. The criterion for marginal stability to overturning convection is that the vertical lagrangian derivative of the density be zero (i.e. that the Brunt-V\"ais\"al\"a frequency vanish). To be useful as a stability criterion this requires further information about how a vertical displacement of a fluid element is to be effected.  If the fluid element is displaced isentropically, as is typically assumed, then the marginally stable state is adiabatic. Thus
one commonly expects that efficient convection will relax a layer to an adiabatic state. Here, however, vertical displacements of fluid elements are not isentropic; heat is constantly being supplied to the fluid elements by dissipative heating driven by the MRI. Furthermore, the source of free energy, here is not the stratification as is the case in straight thermal convection, but the rotational shear. Thus there is no {\it a priori} reason why the mean stratification should relax to an adiabatic state.

Finally, we comment on our choice of thermal/mechanical boundary conditions in the vertical. As mentioned in \S 2 we adopt radiative, impenetrable stress-free boundaries. The stress-free part is so that the regions external to the computational domain exert no tangential stresses on the interior. This is a reasonable request that the boundary be tangentially neutral. The radiative, impenetrable part, specially in view of the recent work of \citet{Gressel13} requires further consideration. Gressel compares two cases, one similar to the one considered here and one in which the vertical boundary conditions are open. Not surprisingly in this second case the layer puffs up and the convective motions are greatly reduced. He then concludes that the convective state is to a large extent an artifact of the boundary conditions, and that once the more ``natural" open boundary conditions are implemented the convection all but disappears. We feel that this is not quite right. 

Clearly, assuming that there is a horizontal surface (two, actually) on which the vertical velocity vanishes and radiative boundary conditions are applied is an idealization introduced, to some extent, for numerical simplicity. However, it also has a physical meaning. In a realistic situation, there should be no physical boundary, but rather a region, and depending on the details, possibly a very thin region, where the plasma changes from optically thick to optically thin. In general, this region is characterized by strong gradients and complex physical processes that cause and are caused by the abrupt changes in optical thickness. With suitable resources it is possible to model this transition layer to some degree of accuracy. However, from the point of view of the bulk of the convection, the  effect of such a layer is just to provide a source of low entropy that causes the reversal of the buoyancy forces acting on moving fluid parcels. In other words when hot up-flowing fluid reaches the transition layer, it is rapidly cooled off, becomes anti-buoyant and falls back down again. If one were primarily interested in the dynamics of the interior, and the transition regions were thin, one could justify an idealization in which the complexities of the transition regions were replaced by impenetrable radiative boundaries, as was done in the present work. If on the other hand, open boundary conditions are applied  while keeping the thermal diffusivity constant, as was done in Gressell's work, the only source of low entropy is at infinity and the layer puffs up without much convecting. This, however, is like comparing apples and oranges. If  a meaningful comparison is to be made one should compare a layer with an impenetrable boundaries together with radiative  conditions with one with open boundaries  and a mechanism to effect an abrupt transition from very low to very high thermal conductivity, or from very high to very low opacity somewhere in interior. Such an experiment could be quite informative and could probably be undertaken with current resources.

\section{Acknowledgment}
This work was supported in part  by the NSF
sponsored CMSO at the University of Chicago.
GB, AM and PR acknowledge support by  an INAF-PRIN grant. 
The  results in this paper were obtained  using the PRACE Research Infrastructure resource JUQUEEN  at  the J\"ulich Supercomputing Center, and also  resources by XSEDE,  supported by NSF grant number OCI-1053575.

\end{document}